\documentclass[12pt]{article}

\usepackage{times}

\usepackage{amsmath,amssymb}
\usepackage{graphicx}

\newcommand{\gom}{g_\mathrm{om}}

\newcommand{\com}{C_\mathrm{om}}

\newcommand{\pho}{p_\mathrm{HO}}

\newcommand{\dcaption}[1]{\caption{\baselineskip 18pt #1}}

\newenvironment{sciabstract}{%
\begin{quote} \bf}
{\end{quote}}

\title{Optically Measuring Force near the Standard Quantum Limit}

\author
{Sydney Schreppler$^{1\ast}$, Nicolas Spethmann$^{1}$, Nathan Brahms$^{1\dagger}$,\\ Thierry Botter$^{1\ddagger}$, Maryrose Barrios$^{1\S}$, and Dan M. Stamper-Kurn$^{1,2}$\\
\\
\normalsize{$^{1}$Department of Physics, University of California, Berkeley, CA 94720, USA}\\
\normalsize{$^{2}$Materials Sciences Division, Lawrence Berkeley National Laboratory, Berkeley, CA 94720, USA}\\
\\
\normalsize{$^\ast$To whom correspondence should be addressed; E-mail:  sschreppler@berkeley.edu.}\\
\normalsize{$^\dagger$Present address: Arrayent, Redwood City, CA, USA}\\
\normalsize{$^\ddagger$Present address: Jet Propulsion Laboratory, Pasadena, CA, USA}\\
\normalsize{$^\S$Present address: Department of Physics, Harvard University, Cambridge, MA, USA}\\
}

\date{}

\begin{document}

\maketitle

\begin{sciabstract}

The Heisenberg uncertainty principle sets a lower bound on the sensitivity of continuous optical measurements of force. This bound, the standard quantum limit, can only be reached when a mechanical oscillator subjected to the force is unperturbed by its environment, and when measurement imprecision from photon shot-noise is balanced against disturbance from measurement back-action. We apply an external force to the center-of-mass motion of an ultracold atom cloud in a high-finesse optical cavity. The optomechanically transduced response clearly demonstrates the trade-off between measurement imprecision and back-action noise. We achieve a sensitivity that is consistent with theoretical predictions for the quantum limit given the atoms' slight residual thermal disturbance and the photodetection quantum efficiency, and is a factor of 4 above the absolute standard quantum limit.

\end{sciabstract}

Several decades ago, mounting efforts to detect directly the gravitational radiation produced by distant astrophysical sources prompted investigation of the limits of such a measurement imposed by the quantum-mechanical properties of the detectors \cite{Braginsky:1975, Caves:PRL:1980, Caves:RMP:1980}. The predicted limit has remained a theoretical result, as no experiments to date have operated in regimes that would encounter such a measurement boundary. Far from their mechanical quantum ground states, early gravitational wave detectors \cite{Thorne:RMP:1980} and subsequent table-top force-measurement systems \cite{Cohadon:1999, Tittonen:1999} were dominated by thermal noise that masked the contributions of detection uncertainty due to the measurements themselves. Additionally, the problem of technical sources of optical noise emerging at high measurement strength obscured the effect of measurement back-action in macroscopic systems subject to strong optical probing \cite{Tittonen:1999}, delaying until recently the observation of such back-action noise \cite{Murch:2008,Purdy:2013}. A purely quantum limit to force-measurement imprecision occurs when thermal noise has been reduced to the level of zero-point fluctuations. This ``standard quantum limit" (SQL) appears when the noise introduced by continuous measurement is carefully balanced with the statistical fluctuations of the measurement outcome. In the past few years, a wide array of platforms, having sizes spanning many orders of magnitude, have pushed ever closer to the SQL \cite{Rugar:2004, nichol:193110, Ligo:2009, Moser:2013}. Sensitivity to forces as small as $(390~\text{yN})^2/\text{Hz}$ has been reported \cite{Biercuk:2010}, however these measurements are still at best six to eight orders of magnitude less sensitive than the SQL noise power \cite{Anetsberger:2010,Westphal:PRA:2012}.

\begin{figure}[p]
	\centering
	\includegraphics[width=1\linewidth ]{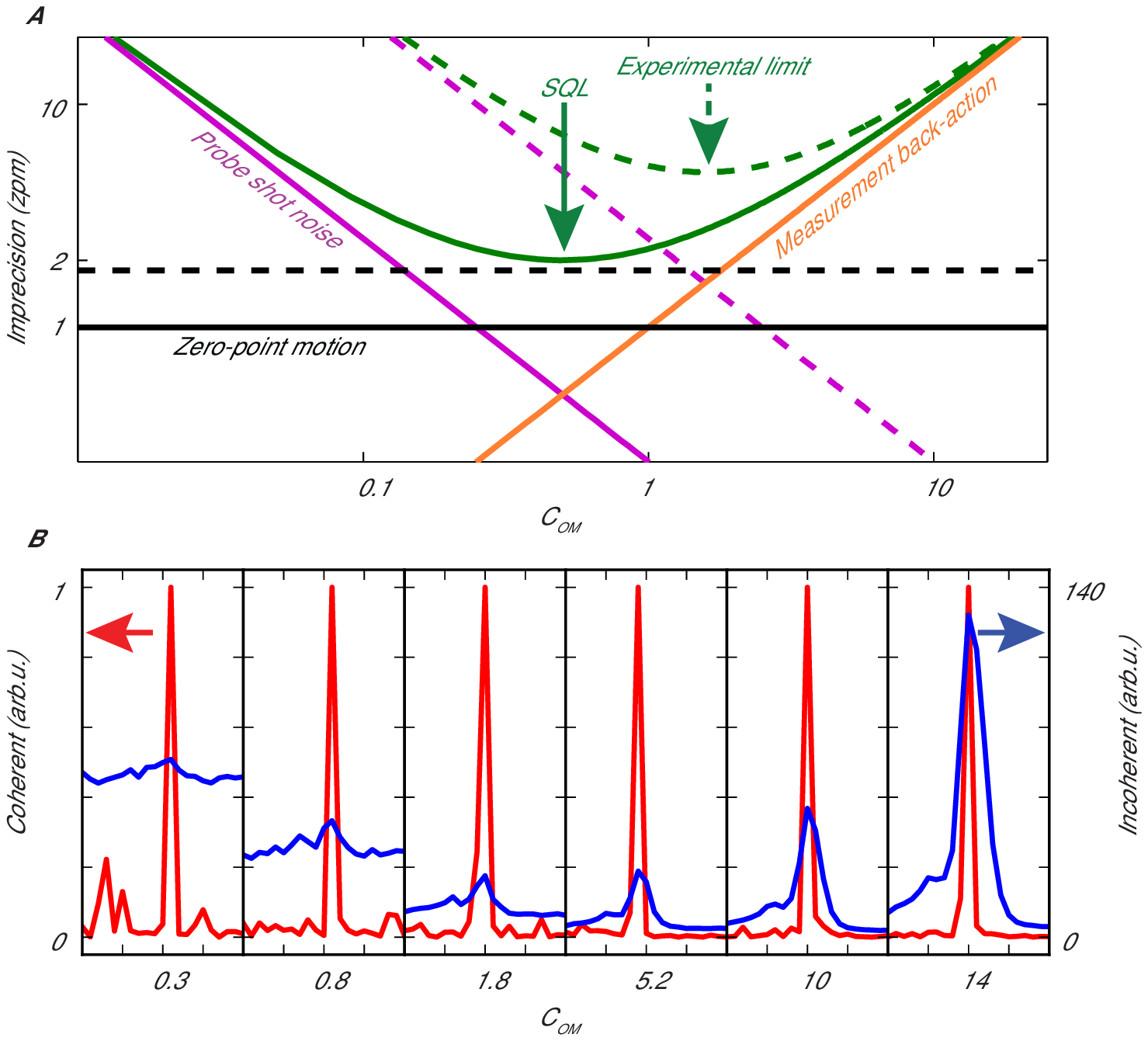}
	\dcaption{Contributions to the standard quantum limit (SQL) of force detection. \textbf{(A)} Theoretical shot-noise imprecision from probe light (purple lines) decreases while measurement back-action imprecision (orange line) increases as $\com$ rises. Together with the zero-point fluctuations (black line), these sum to the total measurement imprecision (green curves). Dashed lines show imprecision bounds corrected for imperfect detection efficiency and small non-zero phonon occupation. Imprecision plotted in units of zero-point motion noise ($\text{zpm}=2\Gamma p_{HO}^2$, as in Eq.\ 1). \textbf{(B)} Experimental data showing the same trend. The incoherent measurement noise (blue curve) is minimized at the mechanical resonance frequency relative to the coherent force response (red curve) at the SQL. Both curves are normalized to the peak coherent response at each value of $\com$.}
	\label{fig:Sensitivity1}
\end{figure}

One way to measure a small force is to apply it to a mechanical harmonic oscillator and then to detect the resulting motion by illuminating the oscillator with light and measuring the phase-shift of the reflected beam. Measurement sensitivity is enhanced by placing the oscillator within a resonant optical cavity \cite{Marquardt:2009}. Quantization of the optical cavity's electromagnetic field (shot noise) sets a lower bound on the imprecision of an optical phase measurement, which varies inversely with measurement strength. In such an optomechanical system, measurement strength can be characterized by the optomechanical cooperativity $\com$, which is proportional to the optical probe power and is understood as the rate at which one gathers information about the oscillator's motion relative to the decay rates of the system. In the regime of low $\com$ the imprecision is mostly due to optical shot noise, while at high $\com$ the contribution from measurement backaction becomes dominant (Fig.~\ref{fig:Sensitivity1}A). The SQL occurs where these two imprecision terms are equivalent.

More quantitatively, the total force measurement imprecision at a given (angular) frequency $\omega = 2\pi \times f$ is
\begin{equation}
S_{FF}(\omega ) = 2 \Gamma \pho^2 \left[\frac{1}{4 \varepsilon \com} \frac{(\omega-\omega_m )^2 +(\Gamma /2)^2}{(\Gamma /2)^2} + (2\nu +1) + \com \right ],
\end{equation}
where $\pho$ is the rms momentum of the harmonic oscillator ground state, $\Gamma$ is the mechanical full-linewidth, $\varepsilon$ is the optical detection efficiency, $\omega _m$ is the mechanical resonance frequency, and $\nu$ is the thermal phonon occupation of the oscillator. The first term in this equation contains the contribution of photon shot-noise, the second the underlying thermal and quantum oscillator fluctuations, and the third the force noise due to measurement back-action. Reduced detection efficiency increases the contribution of shot-noise for all levels of $\com$, raising the minimum sensitivity and moving the force-noise minimum to higher $\com$.

In this work, we characterize a cavity optomechanical system as a force sensor.  We apply a known force to the mechanical element within our system, and measure the motion of this element optically.  The average measured signal calibrates our force sensor.  From the fluctuations of this measurement we then determine the force-measurement noise.  We vary the measurement strength by changing the optical probe power, and realize the quantum limits of our force sensor.

Previous works \cite{Anetsberger:2009, Teufel:NN:2009, Anetsberger:2010, Westphal:PRA:2012} have claimed imprecision below the SQL for \textit{position} detection, taking advantage of the reduction of measurement back-action away from an oscillator's mechanical resonance. For measurement of forces, however, it is clear from Eq.\ 1 that the relative contribution of shot-noise imprecision is minimized at $\omega _m$, allowing for the best sensitivity (green lines in Fig.~\ref{fig:Sensitivity1}A). A large thermal phonon occupation severely reduces sensitivity to applied force on resonance, and indeed this has prevented past measurements of forces at the SQL.
Unlike those previous works, in this report we apply an external force to our oscillator and show that the ultimate \textit{force} measurement imprecision can only be achieved at $\omega _m$.

In our experiments, we apply a calibrated optical-dipole force to a gas of ultracold rubidium atoms, inducing center-of-mass motion of the gas.  We measure this force-induced motion by placing the gas (our forced mechanical oscillator) within a high-finesse Fabry-P\'erot optical cavity and then using a weak probe beam to detect variations in the cavity resonance frequency \cite{Gupta:2007,Brennecke:2008,Purdy:2010}. Before being forced and probed, the mechanical mode is near its motional ground state, with thermal occupation of $\nu = 1.2$ phonons \cite{som}, allowing for unobscured observation of the lower bounds of sensitivity imposed by shot-noise.

We measure the atoms' coherent response to the applied force and incoherent response to probe shot-noise (Fig.~\ref{fig:Sensitivity1}B). When driving on resonance, we clearly observe the three effects discussed above and quantified in Eq.\ 1: shot-noise imprecision decreasing with rising $\com$ and a record of thermal and quantum fluctuations of the oscillator near the mechanical resonance at low $\com$, giving way to much larger fluctuations from back-action noise beyond the sensitivity minimum. Because the probe light is free from technical noise at the frequencies of interest \cite{Brooks:2012}, this back-action is clearly visible and scales as expected with $\com$ \cite{Brahms:PRL:2012}.

\begin{figure}[p]
	\centering
	\includegraphics[width=1\linewidth ]{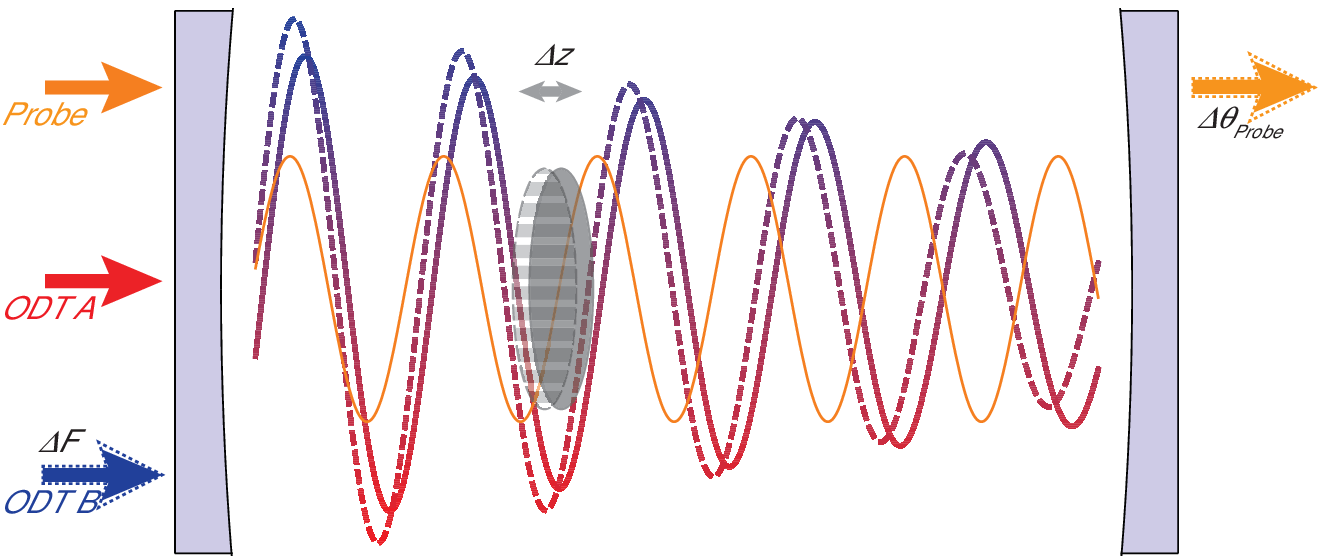}
	\dcaption{Trichromatic lattice potential landscape. Atoms (gray oval) are trapped in a conservative optical-dipole potential created dominantly by two standing-wave light fields with wavelengths of 840 nm and 860 nm (red-blue gradient curve). At the trapping location, the cavity probe light (wavelength 780 nm, orange curve) has a strong intensity gradient, maximizing the position-measurement strength. The 840 nm trap amplitude is modulated (dashed lines) to produce an ac force on the atoms that is optomechanically transduced onto the phase of the probe light. Schematic is not to scale.}
	\label{fig:Schematic}
\end{figure}

We rely on the decoupling of our cold atomic system from its environment to approach the SQL of force sensitivity. Initially, atoms are loaded predominantly into a single site of a trichromatic optical lattice (Fig.~\ref{fig:Schematic}) comprised of two far-off-resonant trapping beams (840 and 860 nm) and one probe beam (780 nm) detuned 24 GHz from atomic resonance. The lattice potential is produced by standing waves of these three lasers, each resonant with a different TEM$_{00}$ (transverse electromagnetic) mode of the optical cavity. The atoms are localized well within the Lamb-Dicke regime, such that the cavity optical mode interacts primarily with their collective center-of-mass mechanical mode \cite{Purdy:2010}. An optical superlattice offers two key advantages for force measurement. First, such a potential landscape allows us to separate spectrally the collective mechanical modes of atoms in neighboring lattice sites, identifying the mode within a single site having resonance frequency $\omega_m$ = 2$\pi \times$ 110 kHz as our oscillator of interest \cite{Botter:PRL:2013}. Second, we can apply a driving force at and around the mechanical resonance frequency of this oscillator by modulating the intensity of one trapping beam while keeping the others fixed. Calculating this applied force requires calibrating both the initial dipole force applied by the trap beams and the modulation index of the drive intensity. The former is achieved by carefully mapping the superlattice potential landscape to determine relative phases and strengths of the component beams at the atoms' location \cite{Botter:PRL:2013}, and is found to be $6.2 \times 10^{-21}$ N per atom. We calibrate the latter with an oscilloscope, finding it to be of order $10^{-8}$.

\begin{figure}[p]
	\centering
	\includegraphics[scale=1]{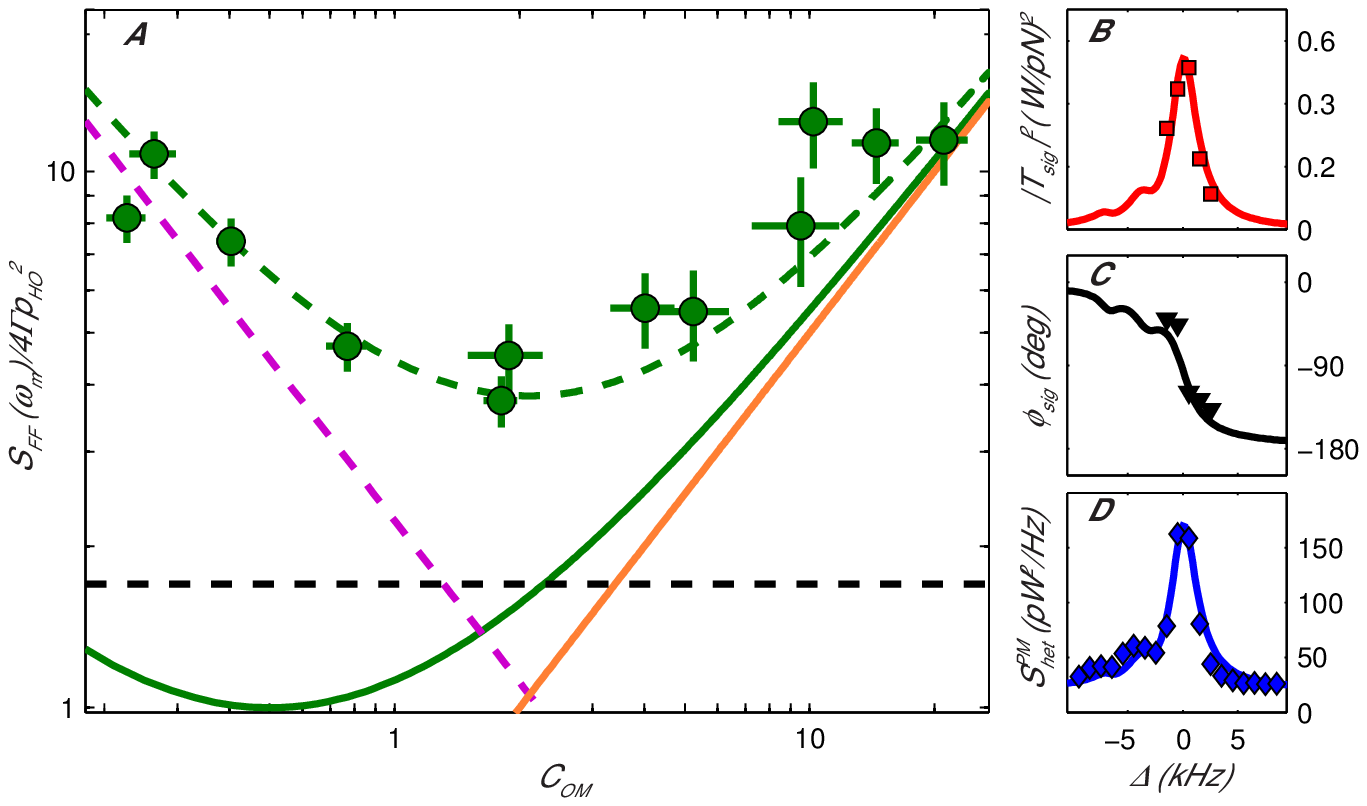}
	\dcaption{Measured sensitivity to applied forces. \textbf{(A)} Sensitivity (ratio of incoherent noise to transduced force) as measured from fits of coherent and incoherent responses, shown over a range of $\com$ and plotted in units of SQL sensitivity (green circles) compared to theory lines for expected precision limits due to probe shot noise (purple line), measurement back action (orange line) and total noise (green curve). Theory lines are shown for perfect (solid) and calibrated (dashed) detection efficiency and phonon occupation. Error bars include contributions from uncertainty of the atom number calibration and of the fits. An example of the simultaneous fits for \textbf{(B)} the squared amplitude $|T_{\text{sig}}|^2$ (red squares) and \textbf{(C)} and phase $\phi_{\text{sig}}$ (black triangles) of the coherent response, and \textbf{(D)} the incoherent noise spectrum $S_{het}^{PM}$ (blue diamonds), plotted against $2\pi\times \Delta = \omega - \omega _m$ for data at $\com$ = 4. See Supplementary materials for definitions of terms. }
	\label{fig:Sensitivity2}
\end{figure}

During separate repetitions of the experiment, we apply the calibrated force at 1 kHz intervals at and around mechanical resonance. In this way, we scan the spectral line shape of the coherent response and are assured of having driven our oscillator within 500 Hz (approximately $\Gamma/6$) of $\omega _m$. The ratio of the complex spectral response to the applied force is the measured susceptibility (Fig.~\ref{fig:Sensitivity2}B,C), which we expect to be proportional to $\chi (\omega) = \left \{ 2 m \omega_m[-(\omega - \omega_m ) - i \Gamma /2]\right \}^{-1}$ for a single mechanical oscillator with mass $m$ and with $\Gamma \ll \omega_m$.

The atoms overlap the probe beam at the point in the standing wave where the light's intensity varies linearly with position (Fig.~\ref{fig:Schematic}). Since the probe light is resonant with the cavity mode, displacement of the atoms maps linearly onto phase fluctuations of the light exiting the cavity, which we detect with a heterodyne receiver. When a force is applied, the driven oscillator response is optomechanically transduced into coherent phase fluctuations of the probe light, while the measurement noise produces incoherent fluctuations.

We tune $\com$ over two orders of magnitude by varying the probe intensity. $\com$ is measured using the optomechanical response to the probe \cite{som}. For each force modulation frequency and at each $\com$, we prepare a new atomic oscillator, address it with a force of constant strength, and optically probe its response. We repeat this process 100 to 200 times per set of measurement parameters and average the results. The heterodyne power spectrum absent the coherent driven response determines the imprecision noise of our measurement. This incoherent shot-noise-driven response gives our total noise spectrum at a given $\com$ (Fig.~\ref{fig:Sensitivity2}D).

We simultaneously fit the coherent driven response and the incoherent noise spectral response of our oscillator at each $\com$ to a Lorentzian resonance condition \cite{som} that includes a primary peak at $\omega_m$ and two secondary peaks slightly red-detuned (Fig.~\ref{fig:Sensitivity2}B-D). These secondary peaks arise from the anharmonicity of our optical superlattice, where atoms excited to higher bands of the trap encounter decreasing level spacing with increasing energy. From each fit we extract the incoherent and coherent response amplitudes as well as $\Gamma$ and $\omega_m$. These parameters, together with the calibrated value of the applied force, determine our force sensitivity on resonance (Fig.~\ref{fig:Sensitivity2}A).

The ratio of the total noise spectrum to the transduced force spectrum gives the spectral sensitivity of Eq. 1, which is minimized when the oscillator is driven resonantly and probed with $\com = 1/2\sqrt{\varepsilon}$, such that $S_{FF}(\omega _m) = 2 \Gamma \pho^2 [(2\nu +1) + 1/\sqrt{\varepsilon}]$. With perfect detection efficiency and oscillator motion limited to zero-point fluctuations, the SQL sensitivity for our oscillator (with $\Gamma = 2\pi \times 3$ kHz and $m = 1.8 \times 10^{-22}$ kg for 1200 $^{87}\textrm{Rb}$ atoms) is $4\Gamma \pho^2 = (21~\text{yN})^2/ \text{Hz}$. Given our system's calibrated photodetection efficiency of 11\%, use of heterodyne rather than homodyne detection (giving $\varepsilon = 0.056$), and average thermal occupation of 1.2 phonons, however, our predicted limit is $(41~\text{yN})^2/\text{Hz}$ and is expected to occur at $\com$ = 2. Our measured force noise is $(42\pm 13~\text{yN})^2/\text{Hz}$, in excellent agreement with theoretical prediction and 4 times higher than the fundamental limit. This corresponds to a measured acceleration sensitivity of $(0.02~g)^2/\textrm{Hz}$, where \textit{g} is the acceleration due to gravity. Both technical limitations of our measurement can be ameliorated straightforwardly: near-infrared photons can be detected with efficiencies exceeding $\varepsilon$ = 0.9 \cite{Marsili:2013} and colder atomic gases can be produced with improved evaporative cooling.

\begin{figure}[p]
	\centering
	\includegraphics[width=1\linewidth ]{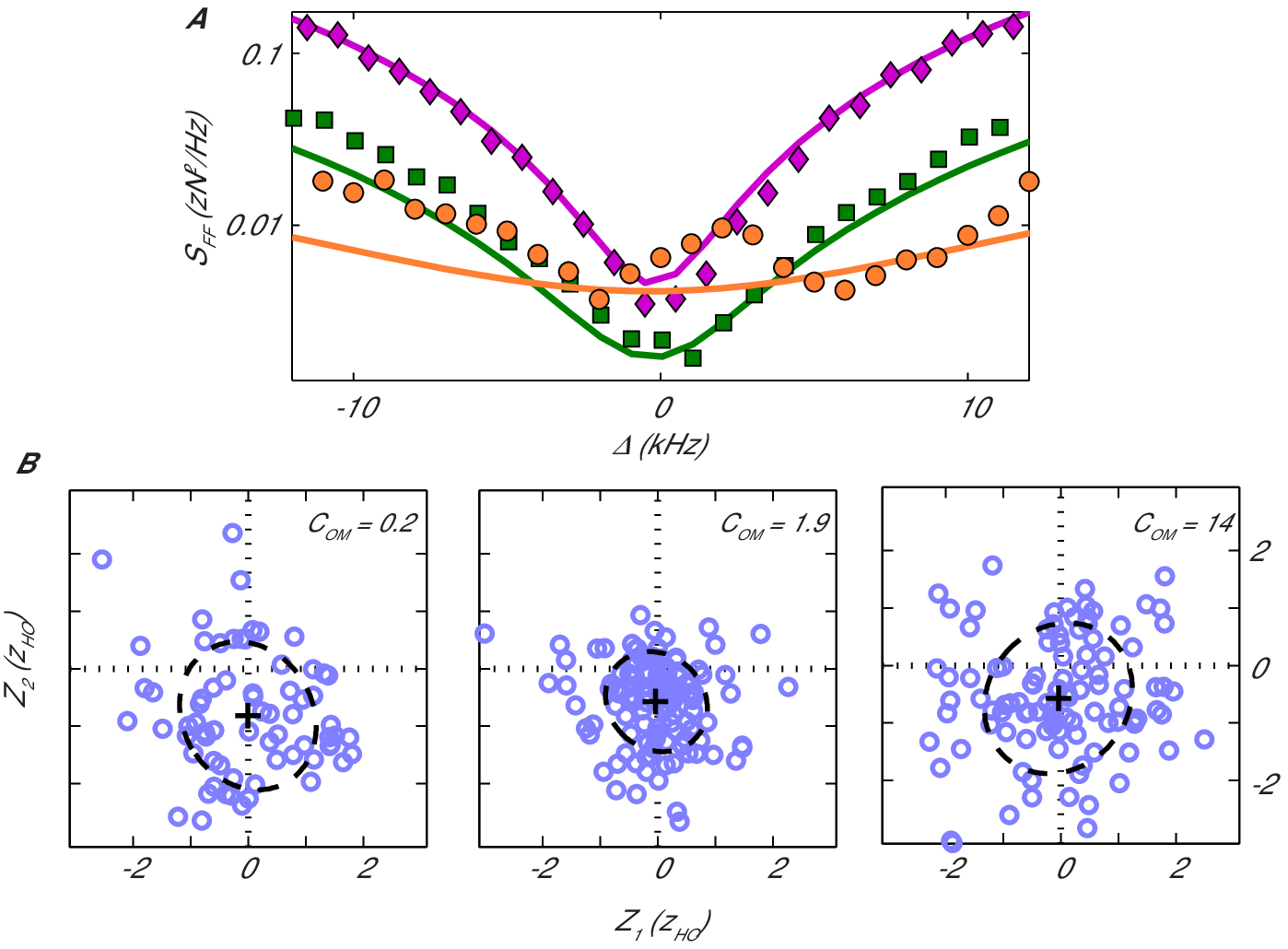}
	\dcaption{Sensitivity spectra and phase space representations. \textbf{(A)} Force spectral density measured at $\com$ = 0.4 (purple diamonds), 1.9 (green squares), and 10 (orange circles), accompanied by theory lines given by Eq.\ 1. \textbf{(B)} Phase space representations of displacements in units of $z_{HO}$ (blue circles) with one-sigma error ellipse (dashed black curves) and average values (black crosses) plotted for $\com$ = 0.2, 1.9, 14 (left to right). Position of black crosses corresponds to average forced displacement.}
	\label{fig:Phase}
\end{figure}

Away from $\omega_m$, force measurement noise diminishes with increasing $\com$ (Fig.~\ref{fig:Sensitivity1}B), because the increasing back-action force noise is still imperceptible due to the reduced mechanical susceptibility. Even for such off-resonant detection, though, the minimum force noise, achieved at $\com>2$, is necessarily greater than the noise limits for forces applied at $\omega _m$ (Fig.~\ref{fig:Phase}A).

Another interpretation of the fundamental limit of force sensitivity states that at the SQL, the total imprecision noise introduced by performing a measurement over the course of one mechanical damping period equals the zero-point fluctuations of the coherent system \cite{Clerk:RMP:2010, Aspelmeyer:arXiv:2013}. This can be seen in Fig.~\ref{fig:Sensitivity1}A, where at the SQL the contributions of shot-noise and measurement back-action sum to one unit of zero-point motion. Therefore, the total imprecision in the detection of forces is at best limited to twice the zero-point noise. Transduction of force to complex optical response involves a necessary intermediate variable, the mechanical element's motion, which can be extracted from our optical data streams. We can therefore observe the quantum-limited imprecision associated with measurement of a displaced mechanical state. Figure~\ref{fig:Phase}B shows these coherent responses measured over 1 ms, \textit{i.e.} averaged over approximately 18 mechanical damping periods.

Uncertainty in the scaled position measurement is related to the force sensitivity by \break $\langle\Delta Z_1\rangle~\langle\Delta Z_2\rangle=|\chi|^2S_{FF}/\tau$, where $\tau$ is the measurement time, and $Z_1$ ($Z_2$) is the real (imaginary) displacement quadrature in units of harmonic oscillator length. Substitution of the minimum $S_{FF}$ produces an uncertainty relation for the imprecision of the state measurement:
\begin{equation}
\langle \Delta Z_1 \rangle ~ \langle \Delta Z_2 \rangle \geq \frac{2}{\tau \Gamma}\left [ (2\nu +1)+1/\sqrt{\varepsilon}\right ].
\end{equation}
The minimum uncertainty condition for this inequality is the quantum limit, consistent with twice the zero-point fluctuations. A covariance matrix of the experimental results gives an error ellipse whose radii quantify the rms imprecision of measurement in the respective quadratures. Near the quantum limit, we observe a spread of points with $\langle \Delta Z_1 \rangle ~ \langle \Delta Z_2 \rangle = 0.8$. As with the measure of force noise imprecision, this value matches our prediction well and is approximately four times the absolute minimum imprecision value.

The achievement of force measurement at the SQL now encourages investigations of sensitivity beyond this level. Several proposed methods that could allow optomechanical systems to overcome this limit rely on noise correlations introduced into the system. Back-action evading measurement techniques \cite{Caves:RMP:1980} have already been demonstrated to reduce classical noise in force detection \cite{Cainard:2007,Hertzberg:2009}, while other proposals call for the coupling of two optical cavity modes \cite{Tsang:PRL:2010,Xu:arXiv:2013,Woolley:2013}, or the measurement of a single quadrature using phase-sensitive quantum non-demolition detection \cite{Caves:RMP:1980, Braginsky:1995:book, Clerk:RMP:2010}. Enhanced force sensitivity and the understanding of its fundamental limits may allow for improved atomic force microscopy, the direct detection of gravitational waves \cite{Ligo:2009}, and measurements of corrections to Newtonian gravity \cite{Geraci:2010}.

\bibliographystyle{science}
\bibliography{refsArxiv}

\begin{thebibliography}{10}
\bibitem{som}
Materials and methods are available as supporting material.
\end{thebibliography}

\noindent
\textbf{Acknowledgements}\\
This work was supported by the AFSOR and NSF. SS was supported by the DoD through the NDSEG Fellowship Program. \\

\clearpage

\clearpage
\newpage

\section{Supplementary Materials}
\setcounter{equation}{0}
\setcounter{figure}{0}
\setcounter{page}{1}
\renewcommand{\theequation}{S\arabic{equation}}
\renewcommand{\thefigure}{S\arabic{figure}}
\renewcommand{\thepage}{\hspace{-1.75in}\roman{page}}

\subsection{Optical Detection}
\label{s:optDet}
We measure our driven and undriven optomechanical spectra using a heterodyne receiver, recording the beat spectrum between our probe light and
a 10-MHz-detuned local oscillator. Depending on $\com$, we average the response over a measurement interval of between 1 and 5 ms (for higher and lower measurement strengths, respectively). Beyond these intervals, the optomechanical response to shot-noise decreases in magnitude and increases in linewidth, showing increased occupation of the anharmonically-spaced trap energy levels (described in Section 1.3).

\subsection{Theoretical Methods}
\label{s:theoMeth}

Here we describe the process by which we derive the expected SQL of force sensitivity for a mechanical oscillator in an optomechanical system.

\noindent
\textbf{Transduced force}

A calibrated external force with spectrum $F(\omega)=F_0\left (\delta_{\omega,\omega_d}+\delta_{\omega,-\omega_d} \right )$ is applied to a mechanical oscillator. It influences the position fluctuations of the mechanical oscillator according to the mechanical susceptibility:
\begin{equation}
	\langle \hat{z}(\omega ) \rangle = \chi (\omega ) F_0,
\end{equation}
where $\chi (\omega )$ is defined as in the text. In the case of linear optomechanical coupling and probe light resonant with the cavity, we expect these position fluctuations to be optomechanically transduced onto the phase fluctuations of the light exiting the cavity \cite{Botter:PRA:2012},
\begin{equation}
	\langle \hat{\zeta }^{PM}(\omega ) \rangle = \frac{\sqrt{2 \com \Gamma}}{z_{HO} \omega _{BW}}F_0\chi (\omega ),
\end{equation}
where $\omega _{BW}$ is the bandwidth of the discrete Fourier transform.

The power in the phase modulation (PM) quadrature, as measured by a heterodyne detector and averaged over a discrete Fourier window, is
\begin{equation}
	\begin{split}
	P^{PM}(\omega )& =\sqrt{\frac{\varepsilon S_{SN}\omega _{BW}^2}{2}}|\langle \hat{\zeta }^{PM} \rangle | \\
	&= \sqrt{\frac{\varepsilon S_{SN} \com \Gamma F_0^2}{z_{HO}^2}} |\chi (\omega ) |
	\end{split}
\end{equation}
where $S_{SN} = P_{LO}\hbar \omega_0$ is the total shot-noise power spectral density (PSD). The shot-noise PSD in the PM quadrature is half this. $P_{LO}$ is the local oscillator power and $\omega _0/2 \pi$ is the probe light frequency. When we normalize by the calibrated force, we recover an expected force transduction of
\begin{equation}
	\begin{split}
	T_{\text{sig}} (\omega ) &= \frac{P^{PM} (\omega ) }{F_0} \\
	&=\sqrt{\frac{\varepsilon S_{SN}\com \Gamma}{ z_{HO}^2}}|\chi (\omega )|.
	\end{split}
\end{equation}

\noindent
\textbf{Noise spectral density}

Absent an external force, the spectral response of the oscillator due to the shot-noise of the probe gives the ``noise" to compare with the ``signal" that is the transduced force. When the probe light is resonant with the cavity, the spectral response in the PM quadrature takes the form $S_{het}^{PM}(\omega ) = \frac{S_{SN}}{2}[1+2\varepsilon n^{PM}(\omega )]$ \cite{Brahms:PRL:2012}, where $n^{PM}(\omega )$ is the photon spectrum inside the cavity given by
\begin{equation}
	n^{PM}(\omega ) = \frac{\com}{2}\frac{\kappa ^2}{\kappa ^2 +\omega ^2}\frac{\Gamma ^2 (2\nu + 1 + \com )}{(\omega - \omega _m)^2+(\Gamma /2)^2}.
\end{equation}
In the resolved sideband regime ($\omega _m \ll \kappa$), the spectrum of interest reduces to
\begin{equation}
	\begin{split}
	S_{het}^{PM} (\omega ) &= \frac{S_{SN}}{2}\left [1+\varepsilon \com \frac{\Gamma ^2 (2\nu +\com +1)}{(\omega - \omega _m)^2+(\Gamma /2)^2}\right ] \\
	&=\frac{S_{SN}}{2}\left [1+4\varepsilon \com (2\nu +\com +1)m^2\omega _m^2\Gamma ^2 |\chi (\omega )|^2\right ].
	\end{split}
\end{equation}

\noindent
\textbf{Force sensitivity}

The ratio of Eq.\ S6 to the squared magnitude of Eq.\ S4 gives the overall sensitivity in Eq.\ 1. The minimum sensitivity occurs on mechanical resonance when $\com =1/(2\sqrt{\varepsilon})$. This gives a quantum-limited sensitivity of
\begin{equation}
S_{FF} = 2\Gamma p_{HO}^2\left [1/\sqrt{\varepsilon}+(2\nu +1)\right ].
\end{equation}
In the case of ideal measurement ($\varepsilon=1$, $\nu=0$), the absolute SQL sensitivity, therefore, is $S_{FF}^{SQL}=4\Gamma p_{HO}^2$.

\noindent
\textbf{Phase space response}

Position imprecision measurements are related to force imprecision measurements via the susceptibility $\chi(\omega)$. Therefore, the variance in position as measured over time $\tau$ can be related to the force sensitivity theory of Eq. 1 by
\begin{equation}
	\begin{split}
	\langle \Delta Z_1 \rangle ~ \langle \Delta Z_2 \rangle z_{HO}^2 &= \langle \Delta F_1 \rangle ~ \langle \Delta F_2 \rangle |\chi (\omega )|^2 \\
	&= \frac{S_{FF}(\omega )}{\tau}|\chi (\omega )|^2 \\
	&= \frac{2 z_{HO}^2}{\Gamma \tau}\left [(2\nu + 1) + 1/\sqrt{\varepsilon} \right ],
	\end{split}
\end{equation}
when $\omega = \omega _m$. In the case of ideal measurement, this reduces to the minimum uncertainty condition of Eq. 2.

\subsection{Data Analysis}
\label{s:data}
Plotted data points in Fig.~\ref{fig:Sensitivity2}A are the result of fit parameters based on heterodyne data that has been normalized by calibrated experimental values. Data points in Fig.~\ref{fig:Sensitivity2}B-D are experimentally-measured values plotted with examples of these fits. In this section we describe the methods of this scaling and fitting. We note that for this data analysis, only the applied force must be calibrated \textit{a priori}.

\noindent
\textbf{Trap anharmonicity}
Though our sinusoidal optical dipole trap is approximately harmonic for low-energy excitations of the gas, atoms occupying higher energy levels experience a relaxed trapping potential (and thus a lower resonance frequency) due to the departure from harmonicity. The difference in resonance frequency between adjacent energy levels is approximately equal to the recoil frequency of the trap light, $\Delta \omega _m = E_R/\hbar = 2\pi \times 3~\text{kHz}$, where $E_R=\hbar ^2 k_t^2/2m$ is the single photon recoil energy for a single $^{87}$Rb atom in a trap of average wavenumber $k_t = (2\pi/850) \text{nm}^{-1}$.

When fitting the observed mechanical spectra of our oscillator, we allow for occupation in the first and second excited states of the oscillator, taking into account the difference in probe-oscillator coupling for different oscillator energy levels. Since our ultimate force sensitivity measurement is performed only on the primary (ground-state) resonance, we then correct the number of atoms in our ground state oscillator based on these scaled fits. We find from these fits that roughly 2-4\% of the atoms occupy the higher energy levels.

\noindent
\textbf{Measurement of cooperativity}

We infer $\com$ for each level of probe power from the fitted height of the resonant response to incoherent probe shot-noise. Solving Eq.\ S6 for $\com$ allows us to determine its value based on the peak height of the incoherent response normalized by shot-noise:
\begin{equation}
\com = -(\nu +1/2)+\sqrt{(\nu +1/2)^2-\frac{2 S_{het}^{PM}(\omega _m)/S_{SN}-1}{4\varepsilon}}.
\end{equation}
The value of $\nu$ is independently calibrated using time-of-flight measurements of the atomic cloud for a particular experimental cycle and is found to equal 1.2 phonons on average. Additionally, we verify this value by comparing the heights of the Stokes and Anti-Stokes sidebands of our undriven spectra \cite{Brahms:PRL:2012}.

Although we expect the measured value of $\com$ to be consistent with the equation
\begin{equation}
\com = \frac{4 n \gom ^2}{\kappa \Gamma},
\end{equation}
where $n$ is the number of probe photons and $\gom$ is the optomechanical coupling, we find that the values obtained from Eq.\ S9 are systematically below those obtained from Eq.\ S10, by about 40$\%$ on average. This may be due to a reduction in $\gom$ from its theoretically predicted value:
\begin{equation}
	g_{om} = \frac{g_0^2}{\Delta _{ca}} k_p N_a z_{HO},
\end{equation}
where $\Delta _{ca}$ is the detuning of the probe light from atomic resonance, $k_p$ is the probe wavenumber, $N_a$ is the number of atoms that comprise the oscillator, and $g_0$ is the single-atom cavity QED coupling. $\Delta_{ca}$ is measured with a high-precision wavemeter, while $\Gamma$ and $z_{HO}$ come from the fits described later in this supplementary material. The coupling $g_0$ takes into account the mode waist and linewidth of the cavity as well as the dipole moment of the D$_2$ rubidium transition. Assuming the atoms are probed with light that is perfectly $\sigma^+$-polarized with respect to their quantization axis and correcting for the finite radial extent of the trapped atomic cloud gives a calculated value of $g_0= 2 \pi \times 13.0$~MHz. However, deviations from perfect alignment of probe polarization would decrease this coupling rate. $N_a$ is measured by observing the average frequency shift of the cavity, $\Delta \omega _c$, due to the presence of atoms:
\begin{equation}
\Delta \omega _c = \frac{N_a g_0^2}{2 \Delta _{ca}}.
\end{equation}
Through post-selection based on measurements of $\Delta \omega _c$, we eliminate all data with atom numbers outside the range $N_a \pm 10\%$. $N_a$ is further corrected to reflect the number of atoms in the spectrally-isolated lattice site of interest by taking into account the calibrated axial extent of the atomic cloud. Approximately 50\% of the atoms measured by $\Delta \omega _c$ occupy the primary site, while the other 50\% occupy sites with resonance frequency at least $2\pi \times 10$~kHz detuned. We then finally correct $N_a$ for trap anharmonicity, as described in the previous section. This process of measurement and calibration of atom number is performed for each experimental run (each value of $\com$). Noting that $N_a$ is measured using the calculated value of $g_0$, we find that the value of $\com$ as determined by Eq. S10 is proportional to $(g_0)^4$. A 15$\%$ reduction in $g_0$, for example, would be sufficient to reduce the calculated value of $\com$ by half.

\noindent
\textbf{Transduced force}

We fit the complex coherent response, measured via heterodyne detection and normalized by the calibrated force we have applied, to a sum of three Lorentzian conditions, each with the following distribution:
\begin{equation}
	T_{\text{sig}}(\omega ) = A_{\text{sig}}e^{i\phi_{\text{sig}}}\sqrt{\frac{(\Gamma /2)^2}{(\omega -\omega _m)^2 + (\Gamma /2)^2}},
\end{equation}
where $\phi_{\text{sig}} = -\arctan \left (\frac{\Gamma/2}{\omega - \omega _m}\right )$ gives the response angle. $A_{\text{sig}}^2$ is our coherent transduced force response in W$^2$/N$^2$. These are plotted for $\com =4$ in Fig.~\ref{fig:Sensitivity2}B-D.

\noindent
\textbf{Noise spectral density}

To quantify the noise of our measurement, we subtract the coherent response used to determine force transduction from the total heterodyne PSD. We average this spectrum with the total PSD of an oscillator with a force applied far off-resonance. In this way we increase our statistics for measuring the noise spectral density. We fit the spectrum to the same Lorentzian condition as used for the coherent response, including an extra term for the incoherent shot-noise:
\begin{equation}
S_{het}^{PM}(\omega ) = A_{NN}\frac{(\Gamma /2)^2}{(\omega -\omega _m)^2 + (\Gamma /2)^2}+S_{SN}/2.
\end{equation}
From the fit (performed simultaneously with the coherent response fit) we extract $\omega _m$, $\Gamma$, and $A_{NN}$, our incoherent noise response. We use $A_{NN}$, the incoherent noise on our force measurement (in W$^2$/Hz) additionally to determine the $\com$ of the system by substituting $A_{NN} +S_{SN}/2 = S_{het}^{PM}(\omega _m)$ into Eq.\ S9.

\noindent
\textbf{Normalizing by the SQL}

We plot in Fig.~\ref{fig:Sensitivity2} measured sensitivity (($A_{NN}+S_{SN}/2)/A_{\text{sig}}^2$) in units of the ideal SQL ($S_{FF}^{SQL}=4 \Gamma p_{HO}^2$) and compare to theory. The values we plot are therefore:
\begin{equation}
	\begin{split}
	S_{FF}(\omega _m)/S_{FF}^{SQL} &= \frac{A_{NN}+S_{SN}/2}{A_{\text{sig}}^2}\frac{1}{4\Gamma p_{HO}^2} \\
	&= \frac{A_{NN}}{A_{\text{sig}}^2}\frac{1}{2\Gamma N_a m_{Rb} \hbar \omega _m},
	\end{split}
\end{equation}
where $m_{Rb}$ is the mass of one $^{87}$Rb atom. Besides the fit amplitudes and the mechanical linewidth and resonance frequency, the only other experimentally calibrated value is $N_a$. Therefore, uncertainty in the value of $S_{FF}(\omega _m)/S_{FF}^{SQL}$ consists of fit errors for the amplitudes, linewidth, and resonance frequency, as well as calibration uncertainty in the atom number. These values are plotted versus the experimentally-determined $\com$, whose uncertainty contains the fit errors for the incoherent response.

\noindent
\textbf{Sensitivity off resonance}

In Fig.~\ref{fig:Phase}A we consider the case of sensitivity to applied force away from mechanical resonance ($S_{FF}(\omega\neq\omega _m)$). To visualize this, we normalize the entire incoherent noise spectral density by the fit to the coherent response:
\begin{equation}
S_{FF}(\omega )=\frac{S_{het}^{PM}(\omega ) [(\omega - \omega _m)^2+(\Gamma /2)^2]}{A_{\text{sig}}^2(\Gamma /2)^2}.
\end{equation}
We compare this measured sensitivity to theory given by Eq.\ 1, plotted as solid lines in the figure.

\noindent
\textbf{Phase space response}

In Fig.~\ref{fig:Phase}B, we show the phase-space statistics of our measured coherent response to applied force. For each $\com$, we choose the measured frequency point in the coherent response spectrum closest in frequency to the resonance frequency obtained from our fits. We plot this point in the complex plane, rotating the response so that the phase angle is given relative to an initial drive phase. Once rotated, we scale both real and imaginary parts to harmonic oscillator units. By substituting Eq.\ S1 into Eq.\ S3, we obtain
\begin{equation}
	P^{PM} = \sqrt{\varepsilon S_{SN} \com \Gamma} \frac{\langle \hat{z} \rangle}{z_{HO}}.
\end{equation}
Therefore, we plot points in phase space with coordinates [$Z_1$,$Z_2$], where
\begin{equation}
	\begin{split}
	&Z_1 = \sqrt{\varepsilon S_{SN} \com \Gamma}P_{re} \\
	&Z_2 = \sqrt{\varepsilon S_{SN} \com \Gamma}P_{im}. \\
	\end{split}
\end{equation}
$P_{re}$ is the real part of the rotated $P^{PM}$ while $P_{im}$ is the imaginary part. $Z_1$ and $Z_2$ are plotted in units of $z_{HO}$.

To determine the statistical variance in the distribution of measurement points, we calculate the covariance matrix of the data and plot the associated error ellipse with 50$\%$ confidence interval. From the eigenvalues of this matrix we extract the minor and major axis lengths of the error ellipse, corresponding to the imprecisions $\langle \Delta Z_1 \rangle$ and $\langle \Delta Z_2 \rangle$.

\end{document}